\documentclass[twocolumn]{aastex631}
\usepackage{amsmath}
\usepackage{graphicx}
\usepackage{CJK}
\usepackage{booktabs}
\usepackage{hyperref}
\usepackage{wasysym}
\usepackage{xcolor}
\usepackage{floatrow}
\newfloatcommand{capbtabbox}{table}[][\FBwidth]
\hypersetup{hidelinks,
	colorlinks=true,
	allcolors=blue,
	pdfstartview=Fit}
\newcommand\sigmarm{${\sigma}_{_{\rm RM}}$}
\newcommand\bpara{$\lvert \langle B_{_\parallel } \rangle \rvert $}

\begin{document}
  \title{Temporal evolution of depolarization and magnetic field of FRB 20201124A}
\shorttitle{Temporal evolution of depolarization and $B$ of FRB 20201124A}

\author[0000-0001-5653-3787]{Wan-Jin Lu}
\affiliation{School of Astronomy and Space Science, Nanjing University, Nanjing 210093, China}

\author[0000-0002-2171-9861]{Zhen-Yin Zhao}
\affiliation{School of Astronomy and Space Science, Nanjing University, Nanjing 210093, China}

\author[0000-0003-4157-7714]{F. Y. Wang}
\affiliation{School of Astronomy and Space Science, Nanjing University, Nanjing 210093, China}
\affiliation{Key Laboratory of Modern Astronomy and Astrophysics (Nanjing University), Ministry of Education, Nanjing 210093, China}
\affiliation{Purple Mountain Observatory, Chinese Academy of Sciences, Nanjing 210023,  China}

\author[0000-0002-7835-8585]{Z. G. Dai}
\affiliation{Department of Astronomy, University of Science and Technology of China, Hefei 230026, China}

\correspondingauthor{Fa-Yin Wang}
\email{fayinwang@nju.edu.cn}

\begin{abstract}

Fast radio bursts (FRBs) are energetic millisecond phenomena in radio band. Polarimetric studies of repeating FRBs indicate that many of these sources occupy extreme and complex magneto-ionized environments. Recently, a frequency-dependent depolarization has been discovered in several repeating FRBs. However, the temporal evolution of polarization properties is limited by the burst rate and observational cadence of telescopes.
In this letter, the temporal evolution of depolarization in repeating FRB 20201124A is explored.
Using the simultaneous variation of rotation measure and dispersion measure, we also measure the strength of a magnetic field parallel to the line-of-sight. The strength ranges from a few $\mu {\rm G}$ to $10^3\ \mu {\rm G}$.  In addition, we find that the evolution of depolarization and magnetic field traces the evolution of rotation measure.
Our result supports that the variation of depolarization, rotation measure and the magnetic field are determined by the same complex magneto-ionized screen surrounding the FRB source. The derived properties of the screen are consistent with the wind and the decretion disk of a massive star.

\end{abstract}


\section{Introduction}\label{sec:intro}

Fast radio bursts (FRBs) are energetic millisecond pulses in the radio band \citep{lori07}.
They show diverse observational properties, including spectral morphology, energy distribution and polarization properties
\citep{Xiao21,zbreview22, review22}.

Recently, observations show that the polarization of FRBs has complex properties. The first repeating FRB source discovered, FRB 20121102A, is highly linear-polarized, while no circular component was found in L-band \citep{121102nat, 121102_AO22}.
The ultra-large rotation measure \citep[RM $\sim 10^5\ {\rm rad}\ {\rm m}^{-2}$,][]{michilli18} implies the source is located in a complex and extreme magneto-ionic environment.
The decreasing behavior of RM also favors a temporal variation scenario \citep{Hilmarsson2021,121102atel}, i.e., an expanding ejecta \citep{Margalit2018,Piro2018,Zhao2021,Zhao2021b,Katz2022,Yang2023}.
FRB 20201124A shows significant temporal and frequency-dependent evolution of polarization properties \citep{xu22}.
It is the first FRB showing RM reversal \citep{binary22}.
FRB 20190520B has been reported to be one of the weirdest repeating FRBs for its dramatic RM reversal and dispersion measure (DM) excess \citep{190520niu,190520dai}.
A magnetar/Be star binary model is proposed to explain the variations and reversal of RM of these two FRBs \citep{binary22}, in which an FRB-generating magnetar orbits a Be-star companion with a
“decretion” disk. In this model, RM variation is caused by the magnetized decretion disk.
Importantly, a decretion disk with a toroidal magnetic field can naturally result in a sign change of RM.
Subsequently, more repeating FRBs have shown RM variations \citep{180916_CHIME,CHIME23,Kumar2023}, implying a dynamical magneto-ionic environment.
However, FRB 20220912A shows a nearly zero local RM in two months, indicating a clean environment \citep{Zhang2023,Feng2023}.
Recently, a repeating FRB source FRB 20200120E discovered in M81 globular cluster also shows a modest RM according to the constraints on the Galactic halo contribution \citep{Bhardwaj2021,200120Enat}.

Since FRB 20201124A was discovered by the Canadian Hydrogen Intensity Mapping Experiment (CHIME) Collaboration \citep{201124atel}, thousands of bursts were detected and well-measured during several active episodes \citep{xu22,Zhoudj22}. Oscillation of both linear and circular polarization components is revealed from the large samples by sensitive monitoring campaign \citep{xu22}.
Some bursts show depolarization that is concurrent with observed RM variation, which can be explained by differential Faraday rotation along different paths \citep{binary22}. \cite{feng22} reported that five active repeating FRBs show frequency-dependent linear polarization fraction, which can be well described by RM scatter. The possible physical origin is the differential Faraday rotation along different paths \citep{Beniamini2022,yangyp22}.
The frequency-dependent linear polarization is also studied in FRB 20180916B and some other repeating FRBs \citep{180916_CHIME,CHIME23}.
For RM-varying FRBs, the frequency-dependent linear polarization should be time-dependent, due to the changes of RM scatter. However, there is no evidence at present.
In this Letter, we report the temporal evolution of depolarization in FRB 20201124A in a daily timescale for the first time.
After the variations of RM and DM are determined, the strength of the magnetic field \bpara\ along the line of sight is also measured. 

This Letter is organized as follows. We introduce the fitting procedure of the spectra in Section~\ref{sec:spec}. The depolarization fit of burst samples and the temporal variation of the \sigmarm\ and \bpara\ are given in Section~\ref{sec:depol}. Discussion and conclusions are shown in Section~\ref{sec:discussion}.

\begin{figure*}[ht]

\centerline{
\hfill
\includegraphics[width=1\linewidth,angle=0]{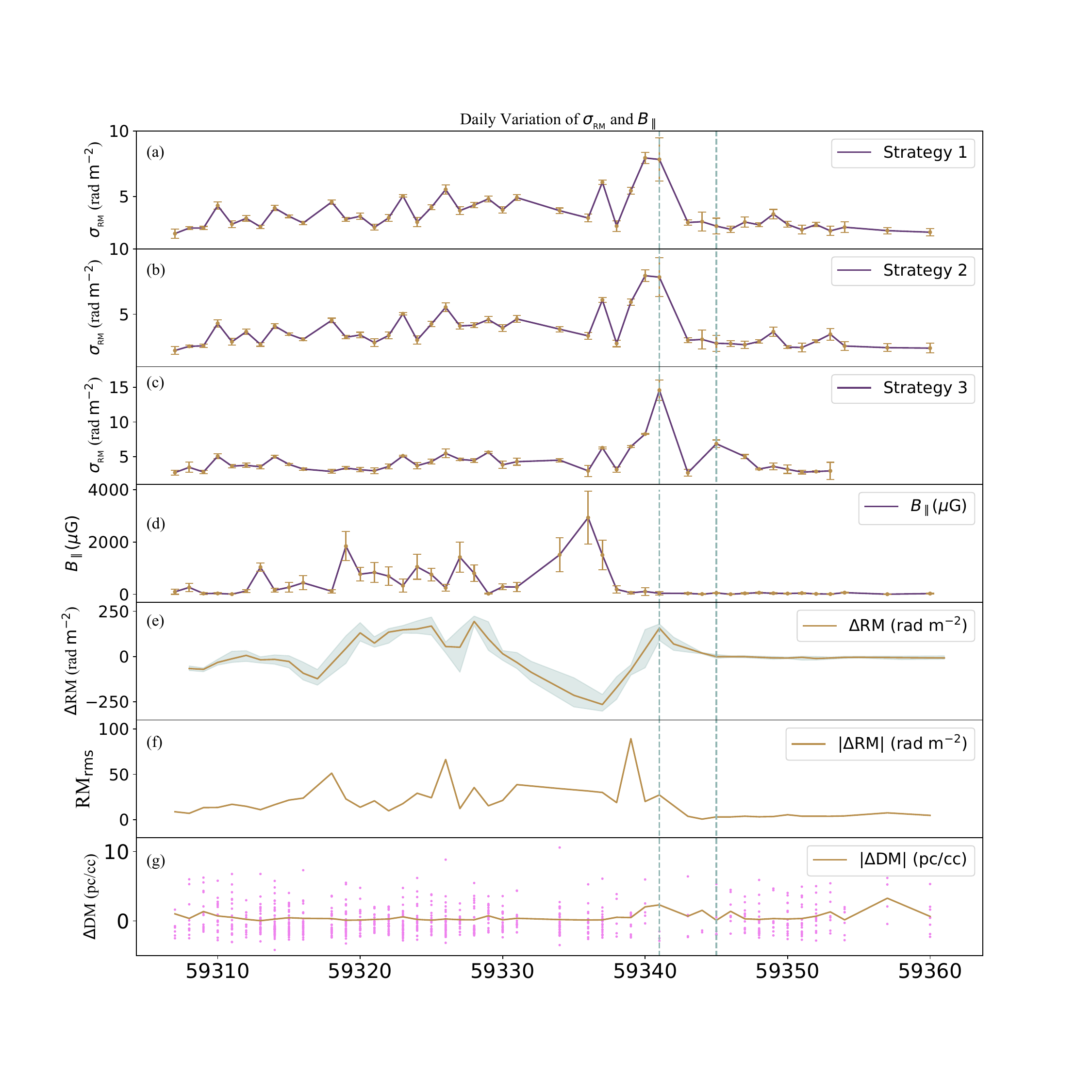}
\hfill
}
\caption{The temporal variation of \sigmarm\ and \bpara\ of FRB 20201124A in a daily timescale. The top three panels show $\sigma_{\rm RM}$ for the three strategies.
The panel (d) shows the variation of $B_{\parallel}$. The panel (e)  gives the RM variation $\Delta$RM=RM$_{\rm obs}$-RM$_c$ and its range defined by the maximum and minimum of the daily RM (the light green region). RM$_c$ is taken to be the average RM value within the active episode, which is 588.06 $\rm rad\ m^{-2}$.
The daily variation of the root-mean-square (rms) of RM is plotted in the panel (f).
The daily variation and the scatter (violet dots) of the $\Delta$DM is shown in the panel (g), similar with the definition of $\Delta$RM.
Two vertical dashed lines are used to show the simultaneous variation trend of \sigmarm\ and $\Delta$RM, which represent MJD = 59341 and 59345, respectively.}

\label{sigmarm_Bpara}
\vspace{-1.5cm}
\end{figure*}

\section{Spectra fitting}\label{sec:spec}
Active FRB repeaters allow us to measure the temporal evolution of the depolarization effect in a short timescale.
However, not all of them have polarimetry measurements.
Limited by the interchannel depolarization caused by the large rotation measure of FRBs and channel frequency of telescopes, repeaters like FRB 20121102A \citep{michilli18} and FRB 20190520B \citep{190520dai} are not suitable for the analysis.
Therefore, we chose the repeating FRB 20201124A monitored by the Five-hundred-meter Aperture Spherical Telescope \citep[FAST,][]{jiang20}. FRB 20201124A shows diverse observational properties, such as the oscillation of the degree of linear polarization (DoL) and temporal evolution of the RM \citep{xu22, jjc22}.

To measure the daily depolarization effect of active FRB repeaters, the DoL and central frequency need to be determined.
\cite{xu22} obtained the DoL and circular polarization component from the Stokes parameters. However, they only estimated the spectral width of each burst by performing the Gaussian fitting and taking the 3$\sigma$ as the effective width owing to the unknown spectra shape of FRBs up-to-date.
Some of the bursts are deviated from the Gaussian function.
Therefore, they did not take the peak value of the Gaussian fitting result as the central frequency.

To fit the central frequency, we took the empirical assumption that the intrinsic spectra of FRBs could be fitted using a Gaussian function. \cite{Zhoudj22} have also discovered a double-peak distribution of central frequency in the observations of FRB 20201124A after May 2021 based on the Gaussian fitting result.
Firstly, we calculated the derivation $\delta f_{\rm i}$ from standard Gaussian spectra for the $i$-th bursts by fitting the Gaussian function
\begin{equation}\label{eq:gauss}
    F(\nu) = F_0 + A \times {\rm exp}[-\frac{(\nu - f_{\rm c})^2}{2\sigma^2}],
\end{equation}
where $\sigma$ is the standard deviation and $f_{\rm c}$ is the central frequency.
With the fitted result above as the initial parameters, we then performed a Markov Chain Monte Carlo (MCMC) fit using the $\tt Python$ package $emcee$ \citep{emcee} to obtain the more accurate result by iteration.

The spectral width $W_{\rm eff}$ is proportional to the fitted standard deviation $\sigma$. The mean value was the fitted central frequency $f_{\rm c}$.
There is also a latent double-peak Gaussian shape similar to the result of \cite{Zhoudj22}.
We did not fit the two peaks of central frequency owing to the selection bias caused by those bursts whose $f_{\rm c}$ sits at the edge of FAST bandwidth.
Therefore, not all the bursts with fitted spectra width and central frequency can be used to study the depolarization effect.
Nevertheless, we determined a series of filter strategies as follows to obtain all the narrow-band samples.
\begin{enumerate}
    \item Set the 3$\sigma$ of the Gaussian function as the equivalent spectral half-width $W_{\rm eff}/2$ and the FAST bandwidth is [1 GHz, 1.5 GHz]. The bursts should satisfy the following criteria:
    1) The central frequency is located in the FAST bandwidth; \\
    2) $W_{\rm eff} > 40\ {\rm MHz}$ and $W_{\rm eff} < 600\ {\rm MHz}$; \\
    3) The lower or upper 60\% equivalent spectral width at least is covered by FAST bandwidth.
    \item Considering the known radio frequency interference, the constraint on the FAST bandwidth is [1.05, 1.45] GHz. The bursts should satisfy the following criteria:
    1) The central frequency is located in the FAST bandwidth;  \\
    2) The $W_{\rm eff}$ satisfies $20\ {\rm MHz} < W_{\rm eff} < 600\ {\rm MHz}$; \\
    3) The lower or upper 85\% equivalent spectral width at least is covered by FAST bandwidth; \\
    4) By artificial identification.  \\
    The spectral width is defined the same as in Strategy 1.
    \item  Set the Full-Width-Half-Maximum (FWHM, i.e.\ $2\sqrt{2\rm ln 2}\sigma$) of the Gaussian function as the equivalent spectral width $W_{\rm eff}$ and the FAST bandwidth is [1 GHz, 1.5 GHz]. Collecting the bursts whose $W_{\rm eff}$ is within the FAST bandwidth.
\end{enumerate}

We collected 761, 808 and 399 bursts out of all 1863 bursts by applying these criteria, respectively. The next step is to fit the depolarization effect from daily bursts.
We sorted the narrow-band bursts above by the barycentric time of arrival at 1.5 GHz \citep{xu22}, from MJD $=$ 59307 to 59360. The fitting function~(Equation~\ref{eq:depol}) from \cite{burn1966}
\begin{equation}\label{eq:depol}
    f_{\rm depol} = 1\ - \exp \left(-2\lambda^4\sigma_{\rm RM}^2\right)
\end{equation}
is used, where $\lambda$ represents the wavelength of the electromagnetic wave in units of $m$ and \sigmarm\ is the standard deviation of the RM.
This function assumes that the FRB source is 100\% linearly polarized originally.
For FRB 20201124A, it has been observed to show frequency-dependent polarization \citep{feng22,201124polkumar22,Effelsberg1124}.
Significant circular bursts were reported by \cite{xu22}, which is considered to be generated by specific radiation mechanism \citep{wwycir23,liucir23,qucir23}.
We then excluded those highly circular polarized bursts from our filtered dataset.
Therefore, Equation~(\ref{eq:depol}) is suitable for studying the temporal evolution of the depolarization effect of FRB 20201124A in L band.

We selected some of the daily depolarization fitting plots in Appendix (see Figure~\ref{extfig}). Though the single-parameter exponential function poorly constrained for some data points, qualitative temporal variation could be distinguished.

Besides, RM is defined as
\begin{equation}
{\rm RM} = 0.81\ {\rm rad \ m^{-2}} \int \frac{\left(B_{\parallel}/\mu {\rm G}\right)n_{\rm e}}{\left(1+z\right)^2}{dl}.
\end{equation}
The average magnetic field parallel to the direction of line-of-sight $\langle B_{_\parallel } \rangle$ could be extracted from the RM measurement.
Furthermore, if the variation of both RM and DM are dominated by the same region (e.g. a plasma scattering screen), the absolute value of $\langle B_{_\parallel } \rangle$ could be estimated as \citep{katz21}
\begin{equation}\label{eq:bpara}
    \lvert \langle B_{_\parallel } \rangle \rvert = 1.23 \ \frac{\left| \Delta{\rm RM}\right| }{ \left| \Delta {\rm DM}\right| }\ \ \mu {\rm G}.
\end{equation}
The $\left| \Delta{\rm RM}\right|$ and $\left| \Delta {\rm DM}\right|$ represent the daily deviations from the average RM and DM value spanning this active episode, respectively.
Equation~(\ref{eq:bpara}) could be applied under the assumption that the variation of RM and DM are contributed by the same outer region of FRB source.
The contribution of the Milky Way and Galactic halo maintains constant at a monthly timescale.

\begin{table*}[htbp]

\begin{center}
\tabcolsep=0.25cm
\caption{Fitting results of \sigmarm\ and \bpara }
\resizebox{0.95\linewidth}{!}{
\hspace*{-3cm}
\begin{tabular}{llllccccccccc}
\hline
\hline
MJD &  \sigmarm  & Counts/$\chi^2$/$p$ & \sigmarm & Counts/$\chi^2$/$p$ & \sigmarm &  Counts/$\chi^2$/$p$  &  \bpara & $\left| \Delta{\rm RM}\right| $ & $\left| \Delta{\rm DM}\right| $ &\\
\hline
& ${\rm rad}\ {\rm m}^{-2}$ &  - & ${\rm rad}\ {\rm m}^{-2}$ &  - & ${\rm rad}\ {\rm m}^{-2}$ &  - & $\mu {\rm G}$ & ${\rm rad}\ {\rm m}^{-2}$ & $\rm cm^{-3}\ pc$ &\\
\hline
\hline

59307	 & $2.2^{+0.4}_{-0.4}$	 & 8/2.95/0.004	 &$2.2^{+0.3}_{-0.3}$	 & 6/1.72/0.054	 &$2.7^{+0.3}_{-0.3}$	 & 4/1.41/0.125	 &95.96	 & 78.76	 & 1.01	 &
 \\
59308	 & $2.6^{+0.1}_{-0.1}$	 & 10/1.51/0.011	 &$2.6^{+0.1}_{-0.1}$	 & 9/1.72/0.016	 &$3.5^{+0.7}_{-0.7}$	 & 3/2.12/0.248	 &263.78	 & 75.57	 & 0.35	 &
 \\
59309	 & $2.6^{+0.1}_{-0.1}$	 & 18/1.47/0.002	 &$2.6^{+0.1}_{-0.1}$	 & 17/1.06/0.002	 &$2.8^{+0.2}_{-0.2}$	 & 11/2.10/0.062	 &30.61	 & 33.55	 & 1.35	 &
 \\
59310	 & $4.3^{+0.3}_{-0.3}$	 & 18/0.27/0.446	 &$4.3^{+0.3}_{-0.3}$	 & 17/0.28/0.449	 &$5.1^{+0.4}_{-0.4}$	 & 6/0.11/0.740	 &32.77	 & 18.70	 & 0.70	 &
 \\
59311	 & $2.9^{+0.3}_{-0.3}$	 & 22/1.47/$<$1e-3	 &$2.9^{+0.3}_{-0.3}$	 & 22/1.42/$<$1e-3	 &$3.6^{+0.3}_{-0.3}$	 & 10/1.01/0.073	 &6.91	 & 2.87	 & 0.51	 &
 \\
59312	 & $3.3^{+0.2}_{-0.2}$	 & 12/1.46/0.013	 &$3.7^{+0.2}_{-0.2}$	 & 15/1.00/0.009	 &$3.8^{+0.3}_{-0.3}$	 & 10/1.10/0.018	 &123.09	 & 22.08	 & 0.22	 &
 \\
59313	 & $2.7^{+0.1}_{-0.1}$	 & 39/2.03/$<$1e-3	 &$2.7^{+0.1}_{-0.1}$	 & 35/1.91/$<$1e-3	 &$3.5^{+0.3}_{-0.3}$	 & 13/0.62/0.026	 &1037.08	 & 18.58	 & 0.02	 &
 \\
59314	 & $4.1^{+0.2}_{-0.2}$	 & 41/1.26/0.022	 &$4.1^{+0.2}_{-0.2}$	 & 44/1.23/0.016	 &$5.0^{+0.2}_{-0.2}$	 & 25/1.12/0.398	 &154.85	 & 32.10	 & 0.26	 &
 \\
59315	 & $3.5^{+0.1}_{-0.1}$	 & 44/1.49/$<$1e-3	 &$3.5^{+0.1}_{-0.1}$	 & 46/1.01/$<$1e-3	 &$3.9^{+0.1}_{-0.1}$	 & 34/0.76/0.031	 &267.91	 & 98.40	 & 0.45	 &
 \\
59316	 & $3.0^{+0.1}_{-0.1}$	 & 30/2.14/$<$1e-3	 &$3.1^{+0.1}_{-0.1}$	 & 36/1.90/$<$1e-3	 &$3.2^{+0.1}_{-0.1}$	 & 14/2.76/0.009	 &440.80	 & 130.89	 & 0.37	 &
 \\
59318	 & $4.6^{+0.2}_{-0.2}$	 & 20/0.84/0.593	 &$4.5^{+0.2}_{-0.2}$	 & 24/0.63/0.971	 &$2.9^{+0.3}_{-0.3}$	 & 18/3.18/$<$1e-3	 &116.99	 & 30.81	 & 0.32	 &
 \\
59319	 & $3.3^{+0.1}_{-0.1}$	 & 28/2.07/$<$1e-3	 &$3.2^{+0.1}_{-0.1}$	 & 32/2.12/$<$1e-3	 &$3.4^{+0.2}_{-0.2}$	 & 11/3.16/0.009	 &1838.71	 & 136.31	 & 0.09	 &
 \\
59320	 & $3.5^{+0.2}_{-0.2}$	 & 20/1.97/$<$1e-3	 &$3.4^{+0.2}_{-0.2}$	 & 21/2.00/$<$1e-3	 &$3.1^{+0.3}_{-0.3}$	 & 11/2.55/0.006	 &770.62	 & 71.77	 & 0.11	 &
 \\
59321	 & $2.6^{+0.2}_{-0.2}$	 & 10/3.49/0.022	 &$2.8^{+0.3}_{-0.3}$	 & 16/3.13/0.001	 &$3.0^{+0.4}_{-0.4}$	 & 11/3.79/0.002	 &835.68	 & 139.26	 & 0.20	 &
 \\
59322	 & $3.4^{+0.3}_{-0.3}$	 & 27/2.29/$<$1e-3	 &$3.4^{+0.3}_{-0.3}$	 & 27/2.23/$<$1e-3	 &$3.6^{+0.4}_{-0.4}$	 & 17/2.63/0.001	 &694.04	 & 148.46	 & 0.26	 &
 \\
59323	 & $5.0^{+0.1}_{-0.1}$	 & 33/0.62/0.396	 &$5.0^{+0.1}_{-0.1}$	 & 38/0.80/0.140	 &$5.1^{+0.1}_{-0.1}$	 & 22/1.11/0.135	 &328.82	 & 155.70	 & 0.58	 &
 \\
59324	 & $3.0^{+0.3}_{-0.3}$	 & 19/3.75/$<$1e-3	 &$3.0^{+0.3}_{-0.3}$	 & 19/3.67/0.001	 &$3.7^{+0.4}_{-0.4}$	 & 13/2.88/0.007	 &1049.43	 & 174.35	 & 0.20	 &
 \\
59325	 & $4.2^{+0.2}_{-0.2}$	 & 17/1.18/0.010	 &$4.3^{+0.2}_{-0.2}$	 & 19/1.08/0.006	 &$4.3^{+0.4}_{-0.4}$	 & 13/1.51/0.013	 &754.52	 & 61.12	 & 0.10	 &
 \\
59326	 & $5.5^{+0.3}_{-0.3}$	 & 30/0.78/0.446	 &$5.5^{+0.3}_{-0.3}$	 & 31/0.81/0.336	 &$5.5^{+0.6}_{-0.6}$	 & 14/0.85/0.196	 &247.46	 & 53.71	 & 0.27	 &
 \\
59327	 & $3.9^{+0.3}_{-0.3}$	 & 19/1.36/0.001	 &$4.1^{+0.2}_{-0.2}$	 & 20/1.10/0.001	 &$4.6^{+0.2}_{-0.2}$	 & 10/1.04/0.011	 &1418.76	 & 192.39	 & 0.17	 &
 \\
59328	 & $4.3^{+0.2}_{-0.2}$	 & 24/0.69/0.035	 &$4.2^{+0.2}_{-0.2}$	 & 29/0.85/0.002	 &$4.4^{+0.3}_{-0.3}$	 & 18/1.06/$<$1e-3	 &803.20	 & 99.27	 & 0.15	 &
 \\
59329	 & $4.8^{+0.2}_{-0.2}$	 & 14/0.40/0.043	 &$4.6^{+0.3}_{-0.3}$	 & 16/0.50/0.029	 &$5.6^{+0.1}_{-0.1}$	 & 8/0.10/0.196	 &24.46	 & 14.57	 & 0.73	 &
 \\
59330	 & $4.0^{+0.2}_{-0.2}$	 & 24/3.15/0.004	 &$4.0^{+0.3}_{-0.3}$	 & 21/1.32/0.001	 &$3.8^{+0.5}_{-0.5}$	 & 9/2.39/0.009	 &286.44	 & 34.96	 & 0.15	 &
 \\
59331	 & $4.9^{+0.2}_{-0.2}$	 & 19/1.01/0.004	 &$4.6^{+0.3}_{-0.3}$	 & 21/1.23/0.002	 &$4.3^{+0.5}_{-0.5}$	 & 10/2.80/0.005	 &278.63	 & 83.22	 & 0.37	 &
 \\
59334	 & $3.9^{+0.2}_{-0.2}$	 & 33/1.74/$<$1e-3	 &$3.9^{+0.2}_{-0.2}$	 & 32/1.74/$<$1e-3	 &$4.5^{+0.2}_{-0.2}$	 & 17/1.79/0.004	 &1509.29	 & 224.80	 & 0.18	 &
 \\
59336	 & $3.4^{+0.3}_{-0.3}$	 & 15/2.05/0.002	 &$3.3^{+0.3}_{-0.3}$	 & 15/1.71/0.004	 &$3.0^{+0.8}_{-0.8}$	 & 3/7.01/0.321	 &2930.00	 & 283.93	 & 0.12	 &
 \\
59337	 & $6.1^{+0.2}_{-0.2}$	 & 26/0.88/0.350	 &$6.1^{+0.2}_{-0.2}$	 & 23/1.00/0.719	 &$6.3^{+0.1}_{-0.1}$	 & 16/1.00/0.402	 &1500.26	 & 173.47	 & 0.14	 &
 \\
59338	 & $2.7^{+0.4}_{-0.4}$	 & 2/0.32/0.616	 &$2.8^{+0.3}_{-0.3}$	 & 3/0.15/0.502	 &$3.1^{+0.4}_{-0.4}$	 & 2/0.11/0.787	 &190.27	 & 79.77	 & 0.52	 &
 \\
59339	 & $5.4^{+0.3}_{-0.3}$	 & 8/0.64/0.064	 &$5.9^{+0.3}_{-0.3}$	 & 9/0.37/0.209	 &$6.5^{+0.1}_{-0.1}$	 & 3/0.23/0.212	 &57.78	 & 22.34	 & 0.48	 &
 \\
59340	 & $7.9^{+0.4}_{-0.4}$	 & 6/0.41/0.113	 &$7.9^{+0.4}_{-0.4}$	 & 6/0.41/0.113	 &$8.3^{+0.1}_{-0.1}$	 & 3/0.04/0.214	 &106.37	 & 174.60	 & 2.02	 &
 \\
59341	 & $7.8^{+1.7}_{-1.7}$	 & 5/2.13/0.932	 &$7.8^{+1.5}_{-1.5}$	 & 6/2.05/0.661	 &$14.6^{+1.5}_{-1.5}$	 & 3/1.89/0.277	 &38.97	 & 73.02	 & 2.30	 &
 \\
59343	 & $3.0^{+0.2}_{-0.2}$	 & 6/1.58/0.164	 &$3.0^{+0.2}_{-0.2}$	 & 7/1.81/0.074	 &$2.7^{+0.5}_{-0.5}$	 & 3/6.48/0.203	 &32.78	 & 17.98	 & 0.67	 &
 \\
59344	 & $3.1^{+0.7}_{-0.7}$	 & 3/3.27/0.217	 &$3.1^{+0.7}_{-0.7}$	 & 3/3.27/0.217	 &-	 & -	  &9.96	 & 12.32	 & 1.52	 &
 \\
59345	 & $2.7^{+0.6}_{-0.6}$	 & 6/4.66/0.031	 &$2.8^{+0.6}_{-0.6}$	 & 6/3.62/0.060	 &$6.9^{+0.5}_{-0.5}$	 & 2/0.10/0.811	 &49.13	 & 3.44	 & 0.09	 &
 \\
59346	 & $2.5^{+0.3}_{-0.3}$	 & 9/2.91/0.022	 &$2.8^{+0.2}_{-0.2}$	 & 10/2.05/0.034	 &-	 & -	  &3.12	 & 3.50	 & 1.38	 &
 \\
59347	 & $3.0^{+0.4}_{-0.4}$	 & 11/0.88/0.157	 &$2.7^{+0.3}_{-0.3}$	 & 16/1.21/0.006	 &$5.0^{+0.3}_{-0.3}$	 & 6/0.20/0.539	 &36.66	 & 8.61	 & 0.29	 &
 \\
59348	 & $2.8^{+0.2}_{-0.2}$	 & 23/0.54/0.015	 &$2.9^{+0.1}_{-0.1}$	 & 22/0.43/0.048	 &$3.2^{+0.1}_{-0.1}$	 & 9/0.32/0.698	 &59.68	 & 10.15	 & 0.21	 &
 \\
59349	 & $3.7^{+0.4}_{-0.4}$	 & 10/0.88/0.062	 &$3.7^{+0.3}_{-0.3}$	 & 11/1.28/0.033	 &$3.6^{+0.5}_{-0.5}$	 & 7/1.80/0.089	 &39.92	 & 10.63	 & 0.33	 &
 \\
59350	 & $2.9^{+0.2}_{-0.2}$	 & 17/1.15/0.010	 &$2.5^{+0.1}_{-0.1}$	 & 17/1.69/0.002	 &$3.2^{+0.6}_{-0.6}$	 & 3/1.07/0.475	 &29.00	 & 6.53	 & 0.28	 &
 \\
59351	 & $2.5^{+0.3}_{-0.3}$	 & 15/2.60/0.027	 &$2.5^{+0.3}_{-0.3}$	 & 15/2.60/0.027	 &$2.8^{+0.3}_{-0.3}$	 & 7/3.30/0.206	 &46.49	 & 12.78	 & 0.34	 &
 \\
59352	 & $2.9^{+0.2}_{-0.2}$	 & 14/0.28/0.015	 &$2.9^{+0.1}_{-0.1}$	 & 15/0.27/0.014	 &$2.9^{+0.1}_{-0.1}$	 & 8/0.13/0.504	 &16.63	 & 9.15	 & 0.68	 &
 \\
59353	 & $2.4^{+0.4}_{-0.4}$	 & 5/1.85/0.040	 &$3.5^{+0.5}_{-0.5}$	 & 7/0.69/0.079	 &$2.9^{+1.2}_{-1.2}$	 & 2/3.84/0.566	 &9.58	 & 10.08	 & 1.29	 &
 \\
59354	 & $2.7^{+0.4}_{-0.4}$	 & 3/0.48/0.297	 &$2.6^{+0.3}_{-0.3}$	 & 6/1.75/0.199	 &-	 & -	  &62.84	 & 7.41	 & 0.15	 &
 \\
59357	 & $2.4^{+0.2}_{-0.2}$	 & 4/0.32/0.288	 &$2.4^{+0.3}_{-0.3}$	 & 5/0.64/0.072	 &-	 & -	  &3.85	 & 10.24	 & 3.27	 &
 \\

\hline
\end{tabular}
}
\label{fitinfo}
\tablecomments{Column 2-3, 4-5, 6-7 are the fitting results, reduced $\chi^2$ and $p$-value by adopting filter strategies 1, 2 and 3 as introduced in Section~\ref{sec:spec}.
Column 8-10 are the line-of-sight magnetic field, $\left| \Delta{\rm RM}\right|$ and $\left| \Delta{\rm DM}\right|$ by applying Equation~\ref{eq:bpara}.
Failed fitting of several datasets are caused by limited burst counts. }
\end{center}
\end{table*}

\section{Temporal variations of \texorpdfstring{\sigmarm}{sigmarm} and \texorpdfstring{\bpara}{Bpara}}\label{sec:depol}

The RM of the repeating FRB 20201124A has been reported to show significant irregular variation throughout the two-month observations performed by FAST \citep{xu22}.
Similar to RM, we obtained the temporal variation of \sigmarm\ with the 1$\sigma$ uncertainty from the daily depolarization fitting result introduced in Section~\ref{sec:spec}. Figure~\ref{sigmarm_Bpara} shows the temporal variation of the depolarization effect for three different selection strategies.
The temporal variations for the three cases are similar, supporting that the selection strategy has little effect on the final result and the temporal evolution of \sigmarm\ is physical.
The \sigmarm\ shows stochastic fluctuation at the early stage and significant amplification during the reversal of $\Delta$RM.
The evolution of \sigmarm\ traces that of $\Delta$RM, which supports they have the same origin.
After the reversal, the variation is quenched, similar to the $\Delta$RM.

Limited by the burst counts, several observations in the later stage could not perform the depolarization fit using strategy 3.
The joint variation trend in Figure~\ref{sigmarm_Bpara} shows that these polarimetric parameters are independent of different selection criteria on narrow-band bursts.
Although some of the depolarization shows underfitting indicated by the  $p$-value, the daily depolarization results are acceptable for two reasons: 1) The fitting function is a single-parameter quartic exponential function; 2) The majority unfitting data points were found before the significant enhancement of \sigmarm\ (i.e. MJD = 59337) and they are clustered near the lower edge of bandwidth. The central frequency of these datapoints is lower than the fitted value because of the limited bandwidth of FAST. Thus, the actual  \sigmarm\ in these days should be lower than the current result, which would support the low and constant evolution at the early stage shown in Figure~\ref{sigmarm_Bpara}. We also test our method with the observational data from \cite{feng22}, \cite{180916_CHIME} and \cite{CHIME23}. The $p$-value of FRB 20190303A, FRB 20190417A and FRB 20180916B are comparable with the validated data points of ours. The low $p$-value may also be caused by other possibilities of depolarization, such as intrinsic property and internal depolarization \citep{180916_CHIME}.

There are several data points misaligned with the RM variation, which could be caused by the limited burst counts within one day and unknown central frequency.
Besides, the fluctuation of \sigmarm\ and RM are quasi-simultaneous with each other. The variation of \sigmarm\ traces the RM variation and supports the physical scenario that the magneto-ionized environment dominates the observational polarization properties of repeating FRB 20201124A, i.e., the depolarization and RM variation.

Taking the assumption that the variations of both RM and DM are contributed by the same region, we calculated the mean observed variation of the RM and DM from the daily data and figured out the variation of \bpara\ using equation~(\ref{eq:bpara}) in the panel (d) of Figure~\ref{sigmarm_Bpara}.
Similar to the $\left| \Delta \rm RM \right|$ introduced in Figure~\ref{sigmarm_Bpara}, the $\left| \Delta \rm DM \right|$ was obtained from the daily deviation of the FAST data-set by maximizing the coherent power of bursts (see \cite{xu22} and references therein).
The strength of magnetic field is from a few $\mu {\rm G}$ to  $10^3\ \mu {\rm G}$, as shown in Figure~\ref{sigmarm_Bpara}. The variation of \bpara\ spans three orders of strength and quickly decreases to nearly zero as the RM is back to a constant, which is also tracked by the variation of the root-mean-square of daily RM (panel (f) in Figure~\ref{sigmarm_Bpara}).
We estimated the uncertainty of \bpara, which could be mainly caused by the small variation of $\Delta$DM. Mild variation of $\Delta$DM with comparable measurement precision would lead to dramatic enhancement of \bpara. The small variation of $\Delta$DM is supported by at least the two following observations. Firstly, the detection of FRB 20201124A requires that the optical depth of free-free absorption should be smaller than unity, which constrains the value of $\Delta$DM is less than unity (see eq.(3) of \citep{xu22}). Secondly, some bursts of this FRB show the characteristic of Faraday conversion \citep{xu22}.  As discussed by \cite{binary22} and \cite{xu22}, a large magnetic field is required for Faraday conversion. Because the value of $\Delta$RM is about a few hundred $ {\rm rad}\ {\rm m}^{-2}$, large magnetic fields require small $\Delta$DM. The error bars in the panel (f) of Figure~\ref{sigmarm_Bpara} describe a temporal evolution trend qualitatively. 

The result suggests an active and turbulent environment that either the value of the magnetic field or the direction is changing. For example, the toroidal configuration in a rotating disk or stellar wind could explain the changing direction of the magnetic field \citep{binary22,zhao23}. We also compare the value of \bpara\ of other FRBs. \cite{Lin2016} proposed that the average magnetic field strengths of FRBs along the line of sight can be derived from the absolute value of RM and DM.
\cite{Wang2020} derived \bpara\ of FRBs by adopting a strong assumption that RM and DM are from the same medium.
The mean value $1.77$ $\mu {\rm G}$ is found, which is significantly lower than our results.
The dataset applied in \cite{Wang2020} only contained a few active FRB repeaters, while the knowledge of Faraday active medium around the FRB source was remarkably expanded at the dawn of 2020s.
Recently, CHIME reported a series of systematic polarimetry measurements of FRB repeaters \citep{CHIME23}.
The line-of-sight magnetic field is comparable with our result ($10^1 \-- 10^3\ \mu $G).
A large value of 3-17 mG is derived for FRB 20121102A \citep{katz21}.

\section{Discussion and Conclusions}\label{sec:discussion}

\begin{figure}
\centerline{
\hfill
\includegraphics[width=1\linewidth,angle=0]{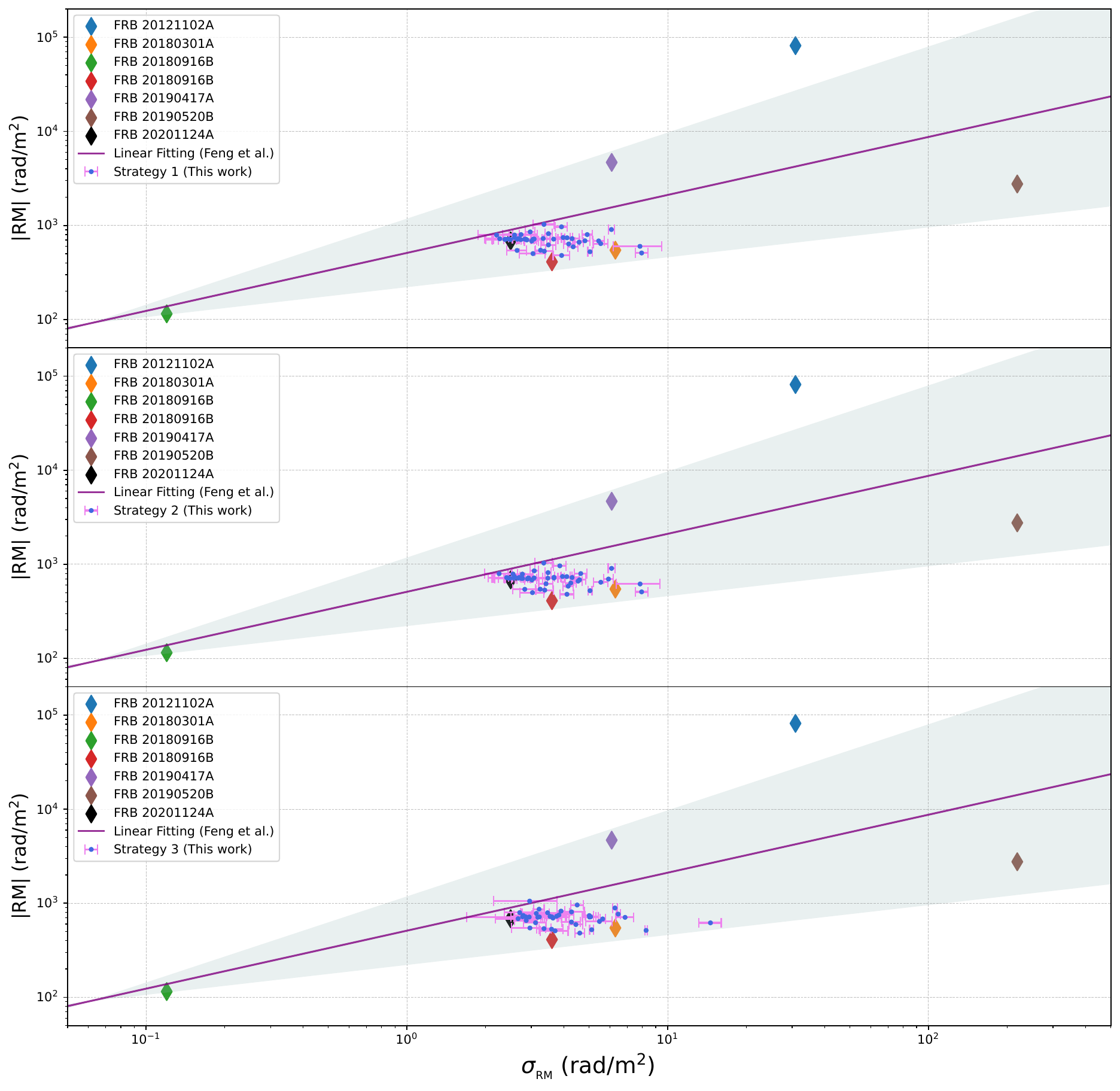}
\hfill
}
\caption{The ${\left| {\rm RM}\right|} $--\sigmarm\ relation. Blue dots with 1$\sigma$ errorbar are the daily result of FRB 20201124A in this work. The red line is the updated fitting linear function with 1$\sigma$ uncertainty and the purple line is from \cite{feng22} with larger uncertainty. }
\label{RM-sigmarm}
\end{figure}

The significant variation of \sigmarm\ and \bpara\ traces the temporal evolution of the local magnetized environment of FRB 20201124A within one active episode.
We updated this daily \sigmarm\ fitting result in the ${\left| {\rm RM}\right|}\ $-- \sigmarm\ relation in \cite{feng22}.
The \sigmarm\ variation is consistent with the 1$\sigma$ uncertainty, as introduced in Figure~\ref{RM-sigmarm}.
The daily RM-\sigmarm\ points from this work actually overweight the FRB 20201124A to provide an inappropriate linear fit for different FRB repeaters.
The scatter of these data points simply shows the variation of FRB 20201124A and is consistent within the 1$\sigma$ uncertainty of \cite{feng22}.

A binary model containing a massive Be star and a magnetar has been applied to explain the stochastic variation of RM and the reversal of the magnetic field \citep{binary22}.
The azimuthal magnetic field distribution in the decretion disk of the Be star provides insight into the RM and magnetic reversal of repeating FRBs such as FRB 20201124A and FRB 20190520B.
Different physical scenarios such as stellar wind and disk have been discussed theoretically \citep{zhao23}.
The temporal variation of \sigmarm\ could constrain different models by applying the depolarization effect caused by multi-path propagation through the plasma screen.

The frequency-dependent relation of linear polarization has been reported and studied in detail among radio pulsars and extragalactic radio sources \citep{burn1966,gardner1966}.
\cite{yangyp22} estimated the value of \sigmarm\ contributed by the screen as
\begin{equation}\label{eq:sigmarm}
    \sigma_{_ {\rm RM}} = 0.81\ {\rm rad}\ {\rm m}^{-2}\ \left( \frac{\sqrt{l_{\rm s}\Delta R}}{1\ {\rm pc}} \right) \left( \frac{\delta (n_e B_{\parallel})_{l_{\rm s}}}{1\ {\rm cm}^{-3}\ \mu {\rm G}} \right).
\end{equation}
The thickness $\Delta R$ and the transverse separation from the line-of-sight $l_{s}$ varies from different models.
$\left( \frac{\delta (n_e B_{\parallel})_{l_{\rm s}}}{1\ {\rm cm}^{-3} \mu {\rm G}} \right)$ represents the fluctuation of electron density and the magnetic field along the line-of-sight.
We compared the contribution of stellar wind and disk, respectively.
For the stellar decretion disk model, we adopted the formula of electron density from Section 2.1 in \cite{zhao23} and assumed the radius at which the electron density maintains half of the value at the stellar surface to be the critical size $r_{d,0.5}$.
For the stellar wind model, we defined the critical size as mentioned in the disk model.

Assuming that the stellar mass $M_*$ and radius $R_*$ to be $8{\rm M}_{\astrosun}$ and $5{\rm R}_{\astrosun}$ respectively, we estimated the transverse separation from the line-of-sight $l_{s}$ when \sigmarm\ varies from 2 to $\sim$8 rad~${\rm m}^{-2}$.
Taking the $\left( \frac{\delta (n_e B_{\parallel})_{l_{\rm s}}}{1\ {\rm cm}^{-3} \mu {\rm G}} \right)\sim\ 10^8 \rm cm^{-3}\ \mu G$ \citep{binary22}, the transparent separation $l_{s}$ in disk model rises from $1.2\times 10^{10}$ to $2\times 10^{11}\ \rm cm$. The variation in stellar wind model is $1.2\times 10^{10}$ to $1.9\times 10^{11}\ \rm cm$.
The result is also comparable with the geometric estimation in \cite{zhao23} and inhomogeneous clumps.

The fluctuation of electron density implies that the massive star in a binary system could originally produce the scattering screen, although these two models are still difficult to distinguish from the temporal variation of \sigmarm. For Be stars, they both contribute to the temporal variations of RM and \sigmarm.
The periastron passage of the magnetar can corrupt the structure of the screen dynamically through the spin-down luminosity and tidal forces \citep{Reig2011,binary22}.
The stellar wind would recover before the next passage (i.e. in one orbital period) due to its high velocity ~3000 km/s \citep{Snow1981}. While the radial velocity of the decretion disk is much less than the sound speed \citep{Okazaki2001}.
The Galactic Be binary system PSR B1259-63/LS 2883 has shown significant differences between the DM and RM of
the pulsed emission from one periastron to the next \citep{Johnston2005}. 
These differences indicate that the local properties
of the Be star disk encountered by the pulsar, such as the disk
density and magnetic field, change considerably between periastron
passages \citep{Johnston2005}. The possible reason is the inhomogeneity of the disk caused by the spin-down luminosity and tidal forces of the pulsar.
Monitoring campaigns in these passage stages may reveal more details in the future and give it a chance to apply in repeating FRBs.

In summary, we have investigated the temporal evolution of depolarization and magnetic field parallel to the line-of-sight in repeating FRBs for the first time.  The strength of a magnetic field is from a few to  $10^3 \mu {\rm G}$. The evolution of depolarization and magnetic field traces the evolution of rotation measure, which supports that the variation of depolarization, rotation measure and the magnetic field are determined by the same complex magneto-ionized screen surrounding the FRB source. In the binary scenario, the derived properties of the screen can put constraints on the wind and the decretion disk of massive stars.


\section*{acknowledgements}
We thank the anonymous referee for helpful comments that
were helpful in improving the manuscript. This work was supported by the National Natural Science Foundation of China (grant Nos. 12273009 and 11833003), the National SKA Program of China (grant Nos. 2022SKA0130100 and 2020SKA0120300), and
the China Manned Spaced Project (CMS-CSST-2021-A12). This work made use of data
from FAST, a Chinese national mega-science facility built and
operated by the National Astronomical Observatories, Chinese
Academy of Sciences.

\bibliographystyle{aasjournal}
\bibliography{ref}

\section{Appendix}

\begin{figure*}[h]
\centerline{
\hfill
\includegraphics[width=1\linewidth,angle=0]{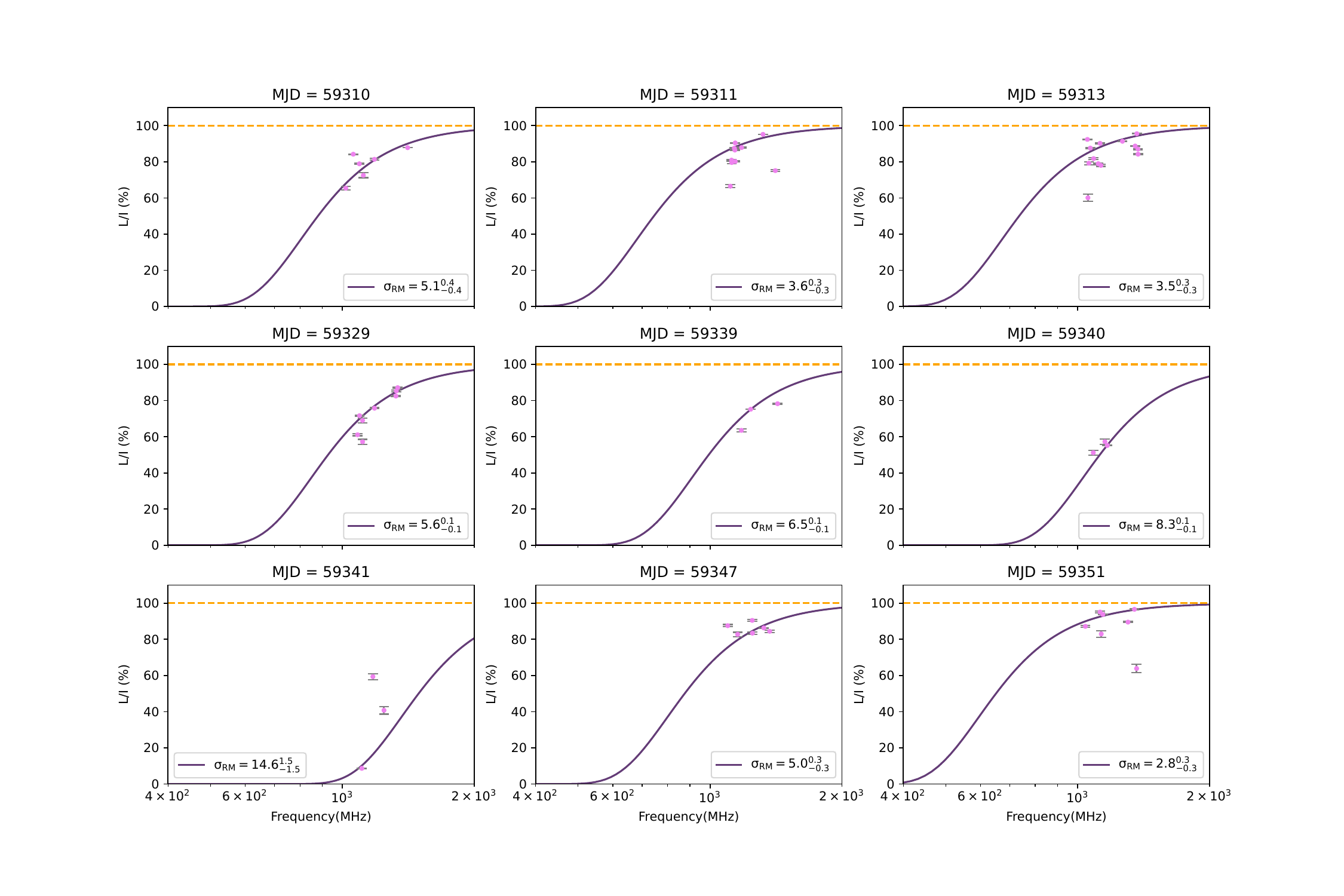}
\hfill
}
\caption{The exemplified daily depolarization fitting plots filtered by strategy 3 (see Section~\ref{sec:spec}. Orange dashed lines represent the 100\% degree of linear polarization. Purple curves show the depolarization fitting results of selected data points (magenta) with error bars.}
\label{extfig}
\end{figure*}

\end{document}